\documentclass[aps,pra,twocolumn]{revtex4}
\usepackage{amssymb}
\usepackage{amsmath}
\usepackage{amsfonts}
\usepackage{bm}
\usepackage{enumerate} 

\begin{document}




\title{Strong entanglement criterion involving momentum weak values}


\author{A. Vald\'{e}s-Hern\'{a}ndez, L. de la Pe\~{n}a, and A. M. Cetto}

\address{Instituto de F\'{i}sica, Universidad Nacional Aut\'{o}noma de M\'{e}xico, A.
P. 20-364, Ciudad de M\'{e}xico, Mexico}

\begin{abstract}
In recent years weak values have been used to explore interesting
quantum features in novel ways. In particular, the real part of the weak value of the momentum operator has been widely studied, mainly in connection with Bohmian trajectories. Here we focus on the imaginary part and its role in relation with the entanglement of a bipartite system. We establish an entanglement criterion based on weak momentum correlations, that allows to discern whether the entanglement is encoded in the amplitude and/or in the
phase of the wave function. Our results throw light on the physical role of the real and imaginary parts of the weak values, and stress the relevance of the latter in the multi-particle scenario.

\end{abstract}
\maketitle





\section{Introduction}
\label{}
The usual operator algebra of quantum mechanics, when applied to the
linear momentum operator, leads to a complex vector in configuration
space composed of a \textit{flux} velocity $\boldsymbol{v}$ and an
\textit{osmotic} or \emph{diffusive} velocity $\boldsymbol{u}$. The
former is widely known as the flow velocity associated with the probability
current, and it is also recognized as the particle velocity field
in Bohmian mechanics \cite{Bohm1}-\cite{Holland}. The 
diffusive velocity, by contrast, has received little attention despite
its intimate connection to distinctive quantum properties \cite{TEQ},
such as the existence of irreducible (quantum) fluctuations and the
nonclassical features related to the so-called quantum (or Bohm) potential.
In fact most of the studies of $\boldsymbol{u}$ have been circumscribed
to the realm of stochastic quantum mechanics or the hydrodynamic (or
classical-like) formulation of (single-particle) quantum mechanics
\cite{BHu}-\cite{VandU}. More specifically, though some analysis has been made of the role of the diffusive velocity in systems composed
of more than one particle \cite{Bohm2,BHu}, a simple and clarifying analysis on its role in bipartite entangled systems has, to our knowledge, never been presented.

Here we carry out such an analysis and show,
first, that the diffusive velocity associated with each of the two
particles plays a prominent role in expressions related to the quantumness
of the system (as in the single-particle case), and more specifically
in connection with entanglement. Notably, correlations involving the
diffusive velocities are obtained that serve as entanglement indicators
and allow us to discern whether the entanglement is encoded in the
probability distribution (\textit{A-entanglement}), and/or in the
phase (\textit{P-entanglement}) of the bipartite wave function. This
discriminating property, together with the fact that such entanglement
criterion involves only bilinear products of the velocities, differs
from the separability criteria for continuous variables that typically
rest on variances and covariance matrices \cite{var1}-\cite{var3},
higher-order moments \cite{hor1,hor2} or entropic functions of global
variables involving canonically conjugate variables \cite{ent1} (for
a recent account of entanglement criteria based on uncertainty relations
see \cite{UR} and references therein).

We further find that the $A$- and $P$- entanglement signalled via
correlations involving the diffusive velocities can be certified in
a natural way by resorting to the \textit{weak values} \cite{Wdef,Wdef2}
associated with the momentum operator. The connection ensues from
the fact that $\boldsymbol{v}$ and $\boldsymbol{u}$ coincide, respectively,
with the real and imaginary parts of the weak value of $\boldsymbol{\hat{p}}$
(with postselection state $\left\vert {\boldsymbol{x}_{1}\boldsymbol{x}_{2}}\right\rangle $),
an observation that has led a number of authors to explore interesting
features of the quantum phenomenon in novel ways.
However, most of the studies so far focus on the real ($\boldsymbol{v}$)
part \cite{Leavens,Wise} and primarily on the (theoretical and experimental)
study of Bohmian trajectories  \cite{Koc}-\cite{MSteinberg}.
In addition to contributing to the discussion of both $\boldsymbol{v}$
\textit{and} $\boldsymbol{u}$ in the weak-value context (see for example \cite{Hiley12}),
here we take a close look at entanglement from such perspective. The
result is an entanglement criterion based on \textit{weak momentum correlations},
valid for any bipartite pure state of continuous variables; the criterion
proposed is strong in the sense that it serves to distinguish between
$A$- and $P$- entanglement. Along our derivations, we delve into the physical meaning of the real and imaginary parts of the weak value of an arbitrary Hermitian operator, and stress their role in the expression for the quantum correlation between a pair of observables.

The Letter is organized as follows. In Section \ref{uandv} we introduce
the flux and diffusive velocities in a bipartite state. Section \ref{Qu}
is devoted to exhibiting the relevance of the diffusive velocities
in the context of the quantum correlation between particle momenta,
thereby bringing to the fore the importance of $\boldsymbol{u}$ in
connection with paradigmatic quantum features. The link between $\boldsymbol{u}$
and quantumness is taken further in Section \ref{uent}, where the
$A$- and $P$- entanglement criteria based on correlations involving
the diffusive velocities are presented. In Section \ref{weaklocal}
we introduce the reader to the weak values, focusing on the role of their real and imaginary parts. In Section \ref{entW} we proceed to
construct the strong entanglement criterion based on weak values of
momentum operators, and also propose a generalization of it. Finally,
we present some conclusions in Section \ref{conc}.

\section{Diffusive velocity and quantumness}

\subsection{Flow and diffusive velocities}

\label{uandv} Consider a two-particle quantum system in a state described
by the wave function 
\begin{equation}
\psi\left(\boldsymbol{x}_{1},\boldsymbol{x}_{2},t\right)=\sqrt{\rho\left(\boldsymbol{x}_{1},\boldsymbol{x}_{2},t\right)}e^{iS(\boldsymbol{x}_{1},\boldsymbol{x}_{2},t)},\label{C18}
\end{equation}
with $S$ a real function and $\rho=\psi^{\ast}\psi$. In what follows
we assume that the system is bounded so that $\psi$ vanishes at infinity.
Let $\boldsymbol{\hat{p}}_{i}=$ $-i\hbar\nabla_{i}$ be the momentum
operator of the $i$-th particle $(i=1,2)$ with mass $m_{i}$. Direct
calculation gives 
\begin{equation}
\boldsymbol{\hat{p}}_{i}\psi=m_{i}\left(\boldsymbol{v}_{i}-i\boldsymbol{u}_{i}\right)\psi,\label{pi}
\end{equation}
where the (real) velocity vectors $\boldsymbol{v}_{i}$ and $\boldsymbol{u}_{i}$
are given, respectively, by 
\begin{equation}
\boldsymbol{v}_{i}=\frac{\hbar}{m_{i}}\nabla_{i}S,\qquad\boldsymbol{u}_{i}=\frac{\hbar}{2m_{i}}\frac{\nabla_{i}\rho}{\rho}.\label{uivi}
\end{equation}

The (quantum) expectation value of $\boldsymbol{\hat{p}}_{i}$, here
denoted by $\left\langle \boldsymbol{\hat{p}}_{i}\right\rangle _{\text{q}}$,
is thus (in what follows all integrations are performed over the entire
configuration space) 
\begin{eqnarray}
\left\langle \boldsymbol{\hat{p}}_{i}\right\rangle _{\text{q}} & = & \int\psi^{\ast}\boldsymbol{\ \hat{p}}_{i}\psi\;d\boldsymbol{x}_{1}d\boldsymbol{x}_{2}\nonumber \\
 & = & \int m_{i}\left(\boldsymbol{v}_{i}-i\boldsymbol{u}_{i}\right)\rho\;d\boldsymbol{x}_{1}d\boldsymbol{x}_{2}=m_{i}\left\langle \boldsymbol{v}_{i}\right\rangle ,\label{pib}
\end{eqnarray}
where $\left\langle \cdot\right\rangle $ (without a subindex) stands
for the mean value of c-numbers (instead of q-numbers), defined as
$\left\langle \cdot\right\rangle =\int\cdot\,\rho\;d\boldsymbol{x}_{1}d\boldsymbol{x}_{2}$.
Notice that in the last equality we took into account that $\left\langle \boldsymbol{u}_{i}\right\rangle =0$,
since $\rho$ vanishes at infinity. The expectation value of $\boldsymbol{\hat{p}}_{i}$
coincides therefore with the mean value of the momentum $m_{i}\boldsymbol{v}_{i},$
defined in terms of the flow velocity $\boldsymbol{v}_{i}.$ This
is the velocity related to the probability current $\boldsymbol{j}_{i}=\rho\boldsymbol{v}_{i}$
that appears in the continuity equation $\tfrac{\partial\rho}{\partial t}+\sum_{i}\nabla_{i}\cdot\boldsymbol{j}_{i}=0.$
In line with Refs. \cite{TEQ,FHiley18}, it represents a mean velocity averaged over an ensemble of individual particles, whereas in Bohmian mechanics \cite{Bohm1,Wise} it is taken as the actual velocity $(d\boldsymbol{x}_{i}/dt)$ of the
$i$-th particle describing the trajectory $\boldsymbol{x}_{i}(t)$.
The diffusive velocity $\boldsymbol{u}_{i}$, by contrast, does not
contribute to $\left\langle \boldsymbol{\hat{p}}_{i}\right\rangle _{\text{q}}$,
and although it appears on an equal footing with $\boldsymbol{v}_{i}$
in Eq. (\ref{pi}), it is normally absent in the usual quantum mechanics
parlance. However, it certainly acquires importance
when dealing with bilinear products of the form $\left\langle \boldsymbol{\hat{p}}_{i}\cdot\boldsymbol{\hat{p}}_{j}\right\rangle _{\text{q}}$,
and particularly in relation with the entanglement between the two
parties. The results below will show that $\boldsymbol{u}$ has a
role of its own, one that allows us to identify this velocity as a
carrier of the quantumness of the system. 
\subsection{Momentum correlations involving $\boldsymbol{u}_{i}$ \label{Qu} }

In order to exhibit the presence of $\boldsymbol{u}_{i}$ in the quantum
features of the bipartite system, we start by resorting to Eq. (\ref{pi})
and write 
\[
\boldsymbol{\hat{p}}_{i}\cdot\boldsymbol{\hat{p}}_{j}\psi=\boldsymbol{\hat{p}}_{i}\cdot\left[m_{j}\left(\boldsymbol{v}_{j}-i\boldsymbol{u}_{j}\right)\psi\right]
\]
\begin{equation}
=m_{i}m_{j}\boldsymbol{v}_{i}\cdot\boldsymbol{v}_{j}\psi+(\pi_{ij}^{uu}+i\pi_{ij}^{uv})\psi,\label{pipj}
\end{equation}
where we have defined 
\begin{eqnarray}
\pi_{ij}^{uu} & = & -m_{i}m_{j}\boldsymbol{u}_{i}\cdot\boldsymbol{u}_{j}-\hbar m_{j}\nabla_{i}\cdot\boldsymbol{u}_{j},\\
\pi_{ij}^{uv} & = & -m_{i}m_{j}\left(\boldsymbol{v}_{j}\cdot\boldsymbol{u}_{i}+\boldsymbol{v}_{i}\cdot\boldsymbol{u}_{j}\right)-\hbar m_{j}\nabla_{i}\cdot\boldsymbol{v}_{j}.
\end{eqnarray}
Notice that since $m_{j}\nabla_{i}\cdot\boldsymbol{u}_{j}=m_{i}\nabla_{j}\cdot\boldsymbol{u}_{i}$,
and $m_{j}\nabla_{i}\cdot\boldsymbol{v}_{j}=m_{i}\nabla_{j}\cdot\boldsymbol{v}_{i}$,
both $\pi_{ij}^{uu}$ and $\pi_{ij}^{uv}$ are symmetric under the
exchange $i\leftrightarrow j$. We thus obtain 
\begin{equation}
\left\langle \boldsymbol{\hat{p}}_{i}\cdot\boldsymbol{\hat{p}}_{j}\right\rangle _{\text{q}}=m_{i}m_{j}\left\langle \boldsymbol{v}_{i}\cdot\boldsymbol{v}_{j}\right\rangle +\left\langle \pi_{ij}^{uu}+i\pi_{ij}^{uv}\right\rangle .\label{<pipj>}
\end{equation}

Now, taking into account that for any bounded vector $\rho\boldsymbol{g}(\boldsymbol{x}_{1},\boldsymbol{x}_{2})$
\begin{eqnarray}
\left\langle \nabla_{i}\cdot\boldsymbol{g}\right\rangle  & = & \int(\nabla_{i}\cdot\boldsymbol{g})\;\rho\;d\boldsymbol{x}_{1}d\boldsymbol{x}_{2}\nonumber \\
 & = & -\int\boldsymbol{g}\cdot(\nabla_{i}\rho)\;d\boldsymbol{x}_{1}d\boldsymbol{x}_{2}\nonumber \\
 & = & -\frac{2m_{i}}{\hbar}\left\langle \boldsymbol{g}\cdot\boldsymbol{u}_{i}\right\rangle ,\label{divu}
\end{eqnarray}
we get (for bounded $\nabla_{i}\rho$ and $\boldsymbol{j}_{i}$, respectively)
\begin{eqnarray}
\left\langle \nabla_{i}\cdot\boldsymbol{u}_{j}\right\rangle  & = & -\frac{2m_{i}}{\hbar}\left\langle \boldsymbol{u}_{i}\cdot\boldsymbol{u}_{j}\right\rangle ,\label{uu}\\
\left\langle \nabla_{i}\cdot\boldsymbol{v}_{j}\right\rangle  & = & -\frac{2m_{i}}{\hbar}\left\langle \boldsymbol{u}_{i}\cdot\boldsymbol{v}_{j}\right\rangle .\label{uv}
\end{eqnarray}
This implies $\left\langle \pi_{ij}^{uu}\right\rangle =m_{i}m_{j}\left\langle \boldsymbol{u}_{i}\cdot\boldsymbol{u}_{j}\right\rangle $
and $\left\langle \pi_{ij}^{uv}\right\rangle =0$, and consequently
from Eqs. (\ref{pib}) and (\ref{<pipj>}),
\begin{equation}
C_{\boldsymbol{\hat{p}}_{i},\boldsymbol{\hat{p}}_{j}}=m_{i}m_{j}C_{\boldsymbol{v}_{i},\boldsymbol{v}_{j}}+m_{i}m_{j}\left\langle \boldsymbol{u}_{i}\cdot\boldsymbol{u}_{j}\right\rangle ,\label{C22}
\end{equation}
with $C_{\boldsymbol{y},\boldsymbol{z}}$ the correlation $C_{\boldsymbol{y},\boldsymbol{z}}=\left\langle \boldsymbol{y}\cdot\boldsymbol{z}\right\rangle -\left\langle \boldsymbol{y}\right\rangle \!\cdot\!\left\langle \boldsymbol{z}\right\rangle $.

Equation (\ref{C22}) shows that the correlation between the diffusive
velocities plays a central role in deviating the quantum correlation
$C_{\boldsymbol{\hat{p}}_{i},\boldsymbol{\hat{p}}_{j}}$ from the
correlation between the flux momenta (or in Bohmian terms, from the
correlation between the actual momenta of the particles). That such
deviation reflects nonclassical features will become clearer below
(see Eq. (\ref{UV})). At this point it can be verified by putting
$i=j$ in the above equations; in particular, Eq. (\ref{C22}) gives
for the quantum momentum dispersion \cite{TEQ}
\begin{equation}
\sigma_{\boldsymbol{\hat{p}}_{i}}^{2}=m_{i}^{2}\sigma_{\boldsymbol{v}_{i}}^{2}+m_{i}^{2}\left\langle \boldsymbol{u}_{i}^{2}\right\rangle ,\label{N25}
\end{equation}
which shows that whilst $\boldsymbol{u}_{i}$ does not contribute
to the expectation value of $\boldsymbol{\hat{p}}_{i}$, it does contribute
to its fluctuations. Moreover, whereas $\sigma_{\boldsymbol{v}_{i}}^{2}$
may vanish, $\left\langle \boldsymbol{u}_{i}^{2}\right\rangle $ is
always \textit{strictly} greater than zero (we are considering bounded
states, so the case $\rho=\,$constant, or rather $\boldsymbol{u}_{i}=0$,
is ruled out from our analysis). In other words, the presence of $\boldsymbol{u}_{i}$
in Eq. (\ref{N25}) reflects the \textit{irreducible} dispersive nature
of the system characteristic of the quantum phenomenon. \footnote{The dispersive nature exhibited by $\boldsymbol{u}_{i}$ can be interpreted
as a causal manifestation of a random behavior of the particle. Further
discussions on $\boldsymbol{u}$ in connection with stochastic quantum
mechanics can be seen in \cite{BHu,Nelson,PeCeSQM}, and within stochastic
electrodynamics in \cite{TEQ}.}

Now for $i=j$, the term $\pi_{ij}^{uu}$ entering in Eq. (\ref{<pipj>})
becomes 
\begin{equation}
\pi_{ii}^{uu}=2m_{i}V_{Qi},\label{pii}
\end{equation}
with 
\begin{equation}
V_{Qi}=-\frac{1}{2}(m_{i}\boldsymbol{u}_{i}^{2}+\hbar\nabla_{i}\cdot\boldsymbol{u}_{i}).\label{VQi}
\end{equation}
Remarkably, $V_{Qi}$ is closely related to the so-called quantum
potential, which lies at the core of Bohmian mechanics. Indeed, the
quantum potential is defined as $V_{Q}=\sum_{k}\tfrac{-\hbar^{2}}{2m_{k}}\tfrac{\nabla_{k}^{2}\sqrt{\rho}}{\sqrt{\rho}}$
(summed over all the particles of the system), which can be rewritten,
using Eqs. (\ref{uivi}) and (\ref{VQi}), as 
\begin{equation}
V_{Q}(\rho)=\sum_{k}V_{Qk}(\boldsymbol{u}_{k}).\label{VQ}
\end{equation}
It is well known that the quantum potential endows the system with its
nonclassical attributes
\cite{Holland}. However, little attention has been paid to the fact
that $V_{Q}$ is directly linked to the diffusive velocities,
as shown in Eq. (\ref{VQi}) (firstly derived in \cite{Hiretal74} in the single-particle problem, using the method outlined above). Notice also that, in line with the above
results, Eq. (\ref{N25}) can alternatively be expressed as $\sigma_{\boldsymbol{\hat{p}}_{i}}^{2}=m_{i}^{2}\sigma_{\boldsymbol{v}_{i}}^{2}+2m_{i}\left\langle V_{Qi}\right\rangle $,
which relates the momentum dispersion with the $i$-th particle's
quantum potential.

Now, direct calculation of $\left\langle \boldsymbol{\hat{x}}_{i}\cdot\boldsymbol{\hat{p}}_{i}-\boldsymbol{\hat{p}}_{i}\cdot\boldsymbol{\hat{x}}_{i}\right\rangle _{\text{q}}$
using Eq. (\ref{divu}), gives
\[
\left\langle \boldsymbol{\hat{x}}_{i}\cdot\boldsymbol{\hat{p}}_{i}-\boldsymbol{\hat{p}}_{i}\cdot\boldsymbol{\hat{x}}_{i}\right\rangle _{\text{q}}=i\hbar\left\langle \nabla_{i}\cdot\boldsymbol{x}_{i}\right\rangle =-2im_{i}\left\langle \boldsymbol{x}_{i}\cdot\boldsymbol{u}_{i}\right\rangle .
\]
This result displays the equivalence (in terms of mean values) between
the fundamental commutator $[\boldsymbol{\hat{x}}_{i},\boldsymbol{\hat{p}}_{i}]\neq0$
and the (nonzero) correlation $\left\langle \boldsymbol{x}_{i}\cdot\boldsymbol{u}_{i}\right\rangle $,
thus revealing an intimate connection between the presence of $\boldsymbol{u}_{i}$
and the far-reaching consequences (as, e.g., the existence of irreductible
fluctuations) of a nonzero fundamental commutator.

The results of this subsection serve to sustain the statement that
$\boldsymbol{u}_{i}$ can be thought of as a kinematic term that bears
the quantumness of the system. In the next section we take this statement
further, by establishing a relation between the diffusive
velocities and the presence of entanglement in state $\psi$. 

\subsection{Role of $\boldsymbol{u}$ in entanglement}

\label{uent}

The state $\psi$ is non-entangled, that is, $\psi\left(\boldsymbol{x}_{1},\boldsymbol{x}_{2},t\right)=\psi_{1}\left(\boldsymbol{x}_{1},t\right)\psi_{2}\left(\boldsymbol{x}_{2},t\right)$
with $\psi_{i}=\sqrt{\rho_{i}}\exp(iS_{i})$ representing the wave
function of subsystem $i$, if and only if: 
\begin{enumerate}[1.]

\item $\rho$ factorizes as $\rho\left(\boldsymbol{x}_{1},\boldsymbol{x}_{2},t\right)=\rho_{1}\left(\boldsymbol{x}_{1},t\right)\rho_{2}\left(\boldsymbol{x}_{2},t\right)$,
and \label{1}

\item $S$ decomposes as $S\left(\boldsymbol{x}_{1},\boldsymbol{x}_{2},t\right)=S_{1}\left(\boldsymbol{x}_{1},t\right)+S_{2}\left(\boldsymbol{x}_{2},t\right)$. \label{2}
\end{enumerate}

For $i\neq j$, we see that condition \ref{1} implies $\nabla_{i}\cdot\boldsymbol{u}_{j}=0$,
whence (using Eq. (\ref{uu})) $\left\langle \boldsymbol{u}_{1}\!\cdot\boldsymbol{u}_{2}\right\rangle =0$.
Analogously, condition \ref{2} implies $\nabla_{i}\cdot\boldsymbol{v}_{j}=0$,
whence (using Eq. (\ref{uv})) $\left\langle \boldsymbol{u}_{1(2)}\!\cdot\boldsymbol{v}_{2(1)}\right\rangle =0$.
This leads to the following entanglement criteria (with $i\neq j$):
\[
\left\langle \boldsymbol{u}_{i}\cdot\boldsymbol{u}_{j}\right\rangle \neq0\Rightarrow \nabla_{i}\cdot\boldsymbol{u}_{j}\neq0\Rightarrow\psi\text{ is A-entangled},
\]
\begin{equation}
\left\langle \boldsymbol{u}_{i}\cdot\boldsymbol{v}_{j}\right\rangle \neq0\Rightarrow \nabla_{i}\cdot\boldsymbol{v}_{j}\neq0\Rightarrow\psi\text{ is P-entangled},\label{UV}
\end{equation}
where the term `A-entangled' indicates that the entanglement is encoded
in the non-factorizability of the \textit{amplitude} $\sqrt{\rho}$,
whereas `P-entangled' means it is encoded in the non-additivity of
the \textit{phase} $S$. With Eqs. (\ref{UV}), the previous observation
that the diffusive velocities typically come up in expressions
that bring to the fore the quantum properties of the system is reinforced,
now in the context of entanglement \textendash considered the most
distinctive quantum feature of composite systems.

Now, returning to Eq. (\ref{C22}), we see that the correlation between
the diffusive velocities contributes to the quantum momentum
correlations, hence any deviation of $C_{\boldsymbol{\hat{p}}_{i},\boldsymbol{\hat{p}}_{j}}$
from the Bohmian momenta correlation constitutes a trace of $A$-entanglement.
However, correlations of the form $\left\langle \boldsymbol{u}_{i}\cdot\boldsymbol{v}_{j}\right\rangle $
do not contribute to $C_{\boldsymbol{\hat{p}}_{i},\boldsymbol{\hat{p}}_{j}}$.
This does not mean that the correlation $C_{\boldsymbol{\hat{p}}_{i},\boldsymbol{\hat{p}}_{j}}$
is insensitive to any P-entanglement present in the correlations $C_{\boldsymbol{v}_{i},\boldsymbol{v}_{j}}$.
Yet this P-entanglement does not modify the quantum correlations with
respect to the (classically expected) correlations between the flux
momenta.

The above considerations invite us to explore whether the \textit{two}
conditions (\ref{UV}) can be brought together into a \textit{single}
quantity endowed with physical meaning, that serves to establish a
strong entanglement criterion in the sense that it would not only
be useful in attesting entanglement, but also in discerning whether
it is encoded in the amplitude and/or in the phase of the wave function.
In the following Section we tackle this problem. 
\section{Weak values of the momentum}

\subsection{Weak values and local mean averages}\label{3.1}

\label{weaklocal} Let us consider an operator $\hat{A}$,
a \textit{preselection} state $\left|{\psi}\right\rangle$, and a
\textit{postselection} state $\left|{\phi}\right\rangle $. Formally,
the corresponding weak value of $\hat{A}$ is a complex number defined
as \cite{Wdef,Wdef2}
\begin{equation}
\langle\hat{A}\rangle_{\text{w}}^{(\psi,\phi)}=\frac{\langle\phi|\hat{A}|\psi\rangle}{\langle\phi|\psi\rangle}.\label{W}
\end{equation}

Operationally, the weak values of (every power of) an Hermitian operator $\hat{A}$ characterize
the relative correction to the detection probability $P_{0}=|\langle\phi|\psi\rangle|^{2}$
due to an intermediate perturbation $\hat{U}_{\alpha}=e^{-i\alpha\hat{A}}$.
Specifically, if the state $\left|{\psi}\right\rangle $ is affected
by the unitary operation $\hat{U}_{\alpha}$, the detection probability
of the postselection state $\left|{\phi}\right\rangle $ is $P_{\alpha}=|\langle\phi|e^{-i\alpha\hat{A}}|\psi\rangle|^{2}$,
whence 
\begin{equation}
\frac{P_{\alpha}}{P_{0}}=\Big|\sum_{n=0}^{\infty}\frac{(-i\alpha)^{n}}{n!}\frac{\langle\phi|\hat{A}^{n}|\psi\rangle}{\langle\phi|\psi\rangle}\Big|^{2}.
\end{equation}
To first order in $\alpha$ (or equivalently for a `weak' perturbation)
the quotient $P_{\alpha}/P_{0}$ goes as $|1-i\alpha\langle\hat{A}\rangle_{\text{w}}^{(\psi,\phi)}|^{2}$,
and the weak value (\ref{W}) completely determines the relative correction
to $P_{0}$ \cite{RMPweak}. 

Physically, the real part of $\langle\hat{A}\rangle_{\text{w}}^{(\psi,\phi)}$, with $\hat A$ an Hermitian operator, can be understood as the `$\phi$-local'
value of the corresponding dynamical variable $A$ in the state $\psi$, when the description is made
in the $\phi$-representation. This can be seen as follows. Given
the state $\left|{\psi(t)}\right\rangle $ and an element $\left|{\phi}\right\rangle $
of an orthonormal basis of the corresponding Hilbert space (in what
follows a continuous one is assumed), the function $\psi(\phi,t)=\langle\phi|\psi\rangle$
gives the state $\psi$ in the $\phi$-representation. Moreover, the
operator $\hat{A}$ in that same representation, $\hat{A}_{\phi}$,
is defined in such a way that $\hat{A}_{\phi}\psi(\phi)=\langle\phi|\hat{A}|\psi\rangle$,
whence 
\begin{equation}
\langle\hat{A}\rangle_{\text{w}}^{(\psi,\phi)}=\frac{\hat{A}_{\phi}\psi(\phi,t)}{\psi(\phi,t)},\label{wv2}
\end{equation}
and the expectation value of $\hat{A}$ in the state $\left|{\psi}\right\rangle $
can be expressed as: 
\begin{eqnarray}
\langle\hat{A}\rangle_{\text{q}} & = & \langle\psi|\hat{A}|\psi\rangle=\int\langle\psi|\phi\rangle\langle\phi|\hat{A}|\psi\rangle d\phi\nonumber \\
 & = & \int\psi^{*}(\phi,t)\hat{A}_{\phi}\psi(\phi,t)d\phi\nonumber \\
 & = & \int\rho(\phi,t)\langle\hat{A}\rangle_{\text{w}}^{(\psi,\phi)}d\phi\nonumber \\
 & = & \langle\text{Re}\,\langle\hat{A}\rangle_{\text{w}}^{(\psi,\phi)}\rangle+i\,\langle\text{Im}\,\langle\hat{A}\rangle_{\text{w}}^{(\psi,\phi)}\rangle,\label{QW}
\end{eqnarray}
where $\rho(\phi,t)=|\psi(\phi,t)|^{2}$ stands for the probability
density function in $\phi$-space. For $\hat{A}$ Hermitian, the last line implies that 
\begin{equation} \label{QW2}
\langle\text{Im}\langle\hat{A}\rangle_{\text{w}}^{(\psi,\phi)}\rangle=0,
\end{equation}
hence Eq. (\ref{QW}) states that the expectation value of $\hat{A}=\hat{A}^{\dagger}$ in the state $\left|{\psi}\right\rangle$
is just the average of $\text{Re}\,\langle\hat{A}\rangle_{\text{w}}^{(\psi,\phi)}$
weighted with the probability distribution $\rho(\phi,t)$. In this
sense, $\text{Re}\,\langle\hat{A}\rangle_{\text{w}}^{(\psi,\phi)}$
plays the role of the $\phi$-local (i.e., defined at each point $\phi$)
value of $A$ in the $\phi$-space. For example, if the postselection
state is chosen as $\left|{\phi}\right\rangle =\left|{\boldsymbol{x}}\right\rangle $,
we have 
\begin{eqnarray}
\langle\hat{A}\rangle_{\text{q}} & = & \int\rho(\boldsymbol{x},t)\langle\hat{A}\rangle_{\text{w}}^{(\psi,\boldsymbol{x})}d\boldsymbol{x}\nonumber \\
 & = & \int Q(\boldsymbol{x},\boldsymbol{p},t)A(\boldsymbol{x},\boldsymbol{p})d\boldsymbol{x}\,d\boldsymbol{p},\label{QWx2}
\end{eqnarray}
with $Q$ an appropriate (pseudo)-probability density function in
phase space, such that $\rho(\boldsymbol{x},t)=\int Q\,d\boldsymbol{p}$.
From Eqs. (\ref{QW}) and (\ref{QWx2}) we get that, up to a term
with vanishing mean value, $\text{Re}\,\langle\hat{A}\rangle_{\text{w}}^{(\psi,\boldsymbol{x})}$
coincides with 
\begin{equation}
\langle A\rangle_{\psi}(\boldsymbol{x},t)=\frac{1}{\rho}\int Q(\boldsymbol{x},\boldsymbol{p},t)A(\boldsymbol{x},\boldsymbol{p})d\boldsymbol{p},\label{local}
\end{equation}
which is no other than the ($\boldsymbol{x}$)-local average of the
variable $A$, obtained when we partially average $A$ over the momentum
space. 

The imaginary part of $\langle\hat{A}\rangle_{\text{w}}^{(\psi,\phi)}$, in its turn, becomes relevant when bilinear expressions, specifically correlations, are considered (see also \cite{Dress12}). Let $\hat A$ and $\hat B$ denote two Hermitian and commuting operators (so that $\hat A\hat B$ is also Hermitian). With the aid of Eqs. (\ref{QW}) and (\ref{QW2}) it can be shown that the quantum correlation between $\hat A$ and $\hat B$ reads (we omit the superindex $(\psi,\phi)$ in the expression for the weak values) 
\begin{eqnarray}
C_{\hat{A},\hat{B}}&=&\langle \hat{A}\hat{B}\rangle_{\textrm{q}} -\langle \hat{A}\rangle_{\textrm{q}} \langle \hat{B}\rangle_{\textrm{q}} \nonumber\\
&=&\langle\text{Re}\,\langle\hat{A}\hat{B}\rangle_{\text{w}}\rangle-
\langle\text{Re}\,\langle\hat{A}\rangle_{\text{w}}\rangle\langle\text{Re}\,\langle\hat{B}\rangle_{\text{w}}\rangle\nonumber
\end{eqnarray}
\begin{equation}\label{QvsW}
=C_{\text{Re}\langle\hat{A}\rangle_{\text{w}},\text{Re}\langle\hat{B}\rangle_{\text{w}}}-C_{\text{Im}\langle\hat{A}\rangle_{\text{w}},\text{Im}\langle\hat{B}\rangle_{\text{w}}}+\text{Re}\,\left\langle C_{\hat A,\hat B}^{\text{w}}\right\rangle,
\end{equation}
where we have defined
\begin{equation}\label{Wcorr}
C_{\hat{A},\hat{B}}^{\text{w}}=\langle \hat{A}\hat{B}\rangle _{\text{w}}-\langle \hat{A}\rangle _{\text{w}}\langle \hat{B}\rangle _{\text{w}}.
\end{equation}
 In what follows we will refer to this quantity as the \textit{weak correlation} (between $\hat A$ and $\hat B$). According to the discussion following Eq. (\ref{QW2}), the first term in Eq. (\ref{QvsW}) can be interpreted as the correlation between the local values of $A$ and $B$. The difference between the latter and the (standard) quantum correlation $C_{\hat{A},\hat{B}}$ is thus determined by the (correlation between the) imaginary parts of $\langle\hat{A}\rangle_{\text{w}}$ and $\langle\hat{B}\rangle_{\text{w}}$, and the (real part of the) weak correlation $C_{\hat{A},\hat{B}}^{\text{w}}$. 
 
The above results serve to enrich the interpretation of the (normalized) cross-Wigner function, studied in relation with the weak value formalism in \cite{Gosson12}. Specifically, its real part plays the role of a quasi-distribution with respect to which the $\phi$-local value $\text{Re}\,\langle\hat{A}\rangle_{\text{w}}^{(\psi,\phi)}$
can be obtained, whereas its imaginary part becomes relevant for the calculation of correlations.


\subsection{Strong entanglement criteria with momentum weak values}

\label{entW} Weak values acquire relevance in our analysis since,
according to Eqs. (\ref{pi}) and (\ref{wv2}), $m_{i}\left(\boldsymbol{v}_{i}-i\boldsymbol{u}_{i}\right)$
is precisely the weak value of the momentum operator of the $i$-th
particle, with preselection state $\left|{\psi(t)}\right\rangle $
and postselection state $\left|{\phi}\right\rangle =\left|{\boldsymbol{x}}\right\rangle =\left|{\boldsymbol{x}_{1}\boldsymbol{x}_{2}}\right\rangle $
\cite{FHiley}. The recognition that the flux velocity is the real
part of $\langle\boldsymbol{\hat{p}}_{i}\rangle_{\text{w}}^{(\psi,\boldsymbol{x})}$ \cite{Leavens,Wise}
has led to the experimental observation of nonlocal effects on Bohmian trajectories using entangled photons \cite{Koc,Mat,BSimon,MSteinberg}, and further proposals
of experimental monitoring of such trajectories \cite{Trav,FHiley}. In its turn, as we have seen above, consideration of the imaginary part of $\langle\boldsymbol{\hat{p}}_{i}\rangle_{\text{w}}^{(\psi,\boldsymbol{x})}$
allows (among other things) for the determination of the quantum potential, and thus for
further studies of Bohmian Mechanics (in the single-particle problem,
$\boldsymbol{u}$ suffices to determine $V_{Q}$; in the multi-particle
problem, the entire set $\{\boldsymbol{u}_{k}\}$ is required). Another
way of monitoring the quantum potential is via the real part of the
weak value of the total kinetic energy, 
\begin{eqnarray}
\text{Re}\left\langle \frac{\boldsymbol{\hat{p}}_{1}^{2}}{2m_{1}}+\frac{\boldsymbol{\hat{p}}_{2}^{2}}{2m_{2}}\right\rangle _{\text{w}}^{(\psi,\boldsymbol{x})}\!\!\!=\frac{1}{2}m_{1}\boldsymbol{v}_{1}^{2}+\frac{1}{2}m_{2}\boldsymbol{v}_{2}^{2}+V_{Q}.\label{pipjW}
\end{eqnarray}

Let us now consider the weak value of the operator $\boldsymbol{\hat{p}}_{i}\cdot\boldsymbol{\hat{p}}_{j}$.
According to Eqs. (\ref{pipj}) and (\ref{wv2}), it is 
\begin{eqnarray}
\langle\boldsymbol{\hat{p}}_{i}\cdot\boldsymbol{\hat{p}}_{j}\rangle_{\text{w}} & = & \frac{\boldsymbol{\hat{p}}_{i}\cdot\boldsymbol{\hat{p}}_{j}\psi}{\psi}\label{pipjW}\\
 & = & m_{i}m_{j}(\boldsymbol{v}_{i}\cdot\boldsymbol{v}_{j}-\boldsymbol{u}_{i}\cdot\boldsymbol{u}_{j})-\hbar m_{j}\nabla_{i}\cdot\boldsymbol{u}_{j}-\nonumber \\
 &  & -im_{i}m_{j}\left(\boldsymbol{v}_{j}\cdot\boldsymbol{u}_{i}+\boldsymbol{v}_{i}\cdot\boldsymbol{u}_{j}\right)-i\hbar m_{j}\nabla_{i}\cdot\boldsymbol{v}_{j}\nonumber \\
 & = & \langle\boldsymbol{\hat{p}}_{i}\rangle_{\text{w}}\cdot\langle\boldsymbol{\hat{p}}_{j}\rangle_{\text{w}}-\hbar m_{j}(\nabla_{i}\cdot\boldsymbol{u}_{j}+i\nabla_{i}\cdot\boldsymbol{v}_{j}),\nonumber 
\end{eqnarray}
so, in line with Eq. (\ref{Wcorr}), the weak correlation between the momenta is given by
\begin{equation}
C_{\boldsymbol{\hat{p}}_{i},\boldsymbol{\hat{p}}_{j}}^{\text{w}}=-\hbar m_{j}(\nabla_{i}\cdot\boldsymbol{u}_{j}+i\nabla_{i}\cdot\boldsymbol{v}_{j}).\label{corrW}
\end{equation}
Substitution of the real part of this expression into Eq. (\ref{QvsW}) gives, using Eq. (\ref{uu}), the correlation (\ref{C22}) as expected. But beyond contributing to $C_{\boldsymbol{\hat{p}}_{i},\boldsymbol{\hat{p}}_{j}}$, the weak correlation $C_{\boldsymbol{\hat{p}}_{i},\boldsymbol{\hat{p}}_{j}}^{\text{w}}$ provides additional information regarding the entanglement. Indeed, in line with Eqs. (\ref{UV}), a strong entanglement criterion can
now be stated as: 
\begin{subequations} \label{critUVW} 
\begin{eqnarray}
\text{Re}\;C_{\boldsymbol{\hat{p}}_{i},\boldsymbol{\hat{p}}_{j}}^{\text{w}}\neq0 & \Rightarrow & \psi(\boldsymbol{x}_{1},\boldsymbol{x}_{2},t)\text{ is A-entangled},\label{crituW}\\
\text{Im}\;C_{\boldsymbol{\hat{p}}_{i},\boldsymbol{\hat{p}}_{j}}^{\text{w}}\neq0 & \Rightarrow & \psi(\boldsymbol{x}_{1},\boldsymbol{x}_{2},t)\text{ is P-entangled}.\label{critvW}
\end{eqnarray}
Thus, $C_{\boldsymbol{\hat{p}}_{i},\boldsymbol{\hat{p}}_{j}}^{\text{w}}$
suffices to determine not only whether $\psi$ is entangled,
but also the type of entanglement involved. According to Eqs. (\ref{critUVW}) and the last paragraphs
in Section \ref{uent}, we see that it is this weak correlation,
and not $C_{\boldsymbol{\hat{p}}_{i},\boldsymbol{\hat{p}}_{j}}$,
what provides information of both types of entanglement on an equal
footing. A proposal to quantify the amount of each kind of entanglement can be seen in \cite{AZ2018}.

In the one-dimensional case, the conditions $\text{Re}\,C_{\boldsymbol{\hat{p}}_{i},\boldsymbol{\hat{p}}_{j}}^{\text{w}}\neq0$
and $\text{Im}\,C_{\boldsymbol{\hat{p}}_{i},\boldsymbol{\hat{p}}_{j}}^{\text{w}}\neq0$
are not only sufficient but also necessary to guarantee the corresponding
type of entanglement. This follows from the fact that in 1D the conditions
$\nabla_{i}\cdot\boldsymbol{u}_{j}=0$ and $\nabla_{i}\cdot\boldsymbol{v}_{j}=0$
become, respectively, $du_{j}/dx_{i}=0$ and $dv_{j}/dx_{i}=0$, which
according to Eq. (\ref{uivi}) lead to $\rho=\rho_{1}(x_{1})\rho_{2}(x_{2})$
and $S=S_{1}(x_{1})+S_{2}(x_{2})$. Consequently, A-entanglement implies
$du_{j}/dx_{i}\neq0$, whereas P-entanglement implies $dv_{j}/dx_{i}\neq0$,
and we are finally led to \end{subequations} \begin{subequations}
\label{critUVW1d} 
\begin{eqnarray}
\text{Re}\;C_{\hat{p}_{i},\hat{p}_{j}}^{\text{w}}\neq0 & \Leftrightarrow & \psi(x_{1},x_{2},t)\text{ is A-entangled},\label{crituW}\\
\text{Im}\;C_{\hat{p}_{i},\hat{p}_{j}}^{\text{w}}\neq0 & \Leftrightarrow & \psi(x_{1},x_{2},t)\text{ is P-entangled}.\label{critvW}
\end{eqnarray}

The structure of the entanglement criteria (\ref{critUVW}) holds
also for other representations and operators, under certain conditions.
Specifically, we can consider operators $\hat{A}$ and $\hat{B}$
representing, respectively, a dynamical variable of particle 1 and
2, and an orthonormal basis $\{\left|{\phi}\right\rangle =|\alpha\beta\rangle=\left|{\alpha}\right\rangle _{1}\otimes|\beta\rangle_{2}\}$
of the bipartite Hilbert space $\mathcal{H}_{1}\otimes\mathcal{H}_{2}$.
In the $\phi$-representation, the state $\left|{\psi(t)}\right\rangle \in\mathcal{H}_{1}\otimes\mathcal{H}_{2}$
is thus described by the wave function $\psi(\alpha,\beta,t)=\sqrt{\rho(\alpha,\beta,t)}e^{iS(\alpha,\beta,t)}$,
and $\hat{A}$ and $\hat{B}$ become represented by local operators
$\hat{A}_{\phi}=\hat{A}_{\alpha}$, and $\hat{B}_{\phi}=\hat{B}_{\beta}$.
If the representation is such that $\hat{\alpha}\left|{\alpha}\right\rangle =\alpha\left|{\alpha}\right\rangle $
with $[\hat{\alpha},\hat{A}]=\pm\,i\hbar$, and $\hat{\beta}\left|{\beta}\right\rangle =\beta\left|{\beta}\right\rangle $
with $[\hat{\beta},\hat{B}]=\pm\,i\hbar$, then $\hat{A}_{\alpha}=\mp\,i\hbar\,\partial/\partial\alpha$,
and $\hat{B}_{\beta}=\mp\,i\hbar\,\partial/\partial\beta$. All this
gives \end{subequations} 
\begin{eqnarray}
\langle\hat{A}\hat{B}\rangle_{\text{w}}^{(\psi,\phi)} & = & \frac{\hat{A}_{\alpha}\hat{B}_{\beta}\psi}{\psi}=\frac{1}{\psi}\hat{A}_{\alpha}\big[\langle\hat{B}\rangle_{\text{w}}^{(\psi,\phi)}\psi\big]\nonumber \\
 & = & \langle\hat{B}\rangle_{\text{w}}^{(\psi,\phi)}\langle\hat{A}\rangle_{\text{w}}^{(\psi,\phi)}\mp\,i\hbar\frac{\partial}{\partial\alpha}\langle\hat{B}\rangle_{\text{w}}^{(\psi,\phi)},
\end{eqnarray}
and consequently 
\begin{eqnarray}
C_{\hat{A},\hat{B}}^{\text{w}} & = & \pm\hbar\frac{\partial}{\partial\alpha}\text{Im}\,\langle\hat{B}\rangle_{\text{w}}^{(\psi,\phi)}\mp i\hbar\frac{\partial}{\partial\alpha}\text{Re}\,\langle\hat{B}\rangle_{\text{w}}^{(\psi,\phi)}.\label{CwAB2}
\end{eqnarray}
Now, direct calculation shows that whenever $\rho(\alpha,\beta,t)=\rho_{1}(\alpha,t)\rho_{2}(\beta,t)$,
then $\partial[\text{Im}\,\langle\hat{B}\rangle_{\text{w}}^{(\psi,\phi)}]/\partial\alpha=0$,
whereas if $S(\alpha,\beta,t)=S_{1}(\alpha,t)+S_{2}(\beta,t)$, then
$\partial[\text{Re}\,\langle\hat{B}\rangle_{\text{w}}^{(\psi,\phi)}]/\partial\alpha=0$.
Gathering results we arrive at Eqs. (\ref{critUVW}), with $\hat{A}$
and $\hat{B}$ instead of $\boldsymbol{\hat{p}}_{1}$ and $\boldsymbol{\hat{p}}_{2}$,
and $\alpha$ and $\beta$ instead of $\boldsymbol{x}_{1}$ and $\boldsymbol{x}_{2}$.

In the general $N$-particle system, the approach just presented can in principle be applied to certify the entanglement between any two subsystems $s_1$ and $s_2$ that result from a given bipartition of the complete system (so that $s_1$ and $s_2$ have, respectively, $N_1$ and $N_2=N-N_1$ particles).\footnote{Criteria for ascertain entanglement between subsets $s_1$ and $s_2$ such that $N_1+N_2<N$ requires to analyze (reduced) density matrices that, in general, correspond to mixed states. This goes beyond the present entanglement criteria, restricted to pure states.}

\section{Concluding remarks}

\label{conc} Weak values of the momentum offer a highly interesting
subject of research that is particularly suitable for the analysis
of paradigmatic quantum features of the system. On one side, the real
part of $\langle\boldsymbol{\hat{p}}_{i}\rangle_{\text{w}}^{(\psi,\boldsymbol{x})}$
allows for the study of quantum (Bohmian) trajectories and their concomitant
nonlocality. On the other hand, as we have emphazised here, the imaginary
part of $\langle\boldsymbol{\hat{p}}_{i}\rangle_{\text{w}}^{(\psi,\boldsymbol{x})}$
is intimately related to characteristic traits of quantumness, and
in particular to entanglement detection. Indeed, the two-velocity
correlations $\left\langle \boldsymbol{u}_{i}\cdot\boldsymbol{u}_{j}\right\rangle $
and $\left\langle \boldsymbol{u}_{i}\cdot\boldsymbol{v}_{j}\right\rangle $,
both involving the diffusive velocity of one of the parties
and referred to mean values of c-numbers, attest to the (A- or P-)
entanglement of the bipartite state, as stated in Eqs. (\ref{UV}).

Interestingly, the usual quantum correlation $C_{\boldsymbol{\hat{p}}_{i},\boldsymbol{\hat{p}}_{j}}$
differs from the Bohmian correlation between the particle momenta
precisely due to the A-entanglement, yet it does not explicitly include
the companion term related to P-entanglement. This asymmetry is overcome
by resorting to the weak-value formalism. Specifically, both types
of entanglement become manifest and can be detected on an equal footing
in the expression for the weak correlation $C_{\boldsymbol{\hat{p}}_{i},\boldsymbol{\hat{p}}_{j}}^{\text{w}}$. More generally, by appeal to pairs of canonically conjugate
operators, both the real and the imaginary parts of the weak values prove to be useful in certifying the entanglement of
the state of the system \textit{and} to determine whether it is encoded
in the wave function's amplitude or phase. 

Besides providing a physical meaning for both, the real and the imaginary part of the weak value of an Hermitian operator, our results point towards the convenience of delving more
deeply into the subject of the imaginary contributions and
their role in the bipartite (and even in the multipartite) case, where
novel entanglement criteria and ways of exploring quantum correlations
may be found.

\section*{Acknowledgements} The authors acknowledge financial support from DGAPA- UNAM through Project PAPIIT IA101918.


\end{document}